\documentclass[11pt]{article}
\usepackage{amssymb}
\usepackage{graphics}
\usepackage{epsfig}
\begin{document}
\title {Electroweak radiative corrections to $e^+e^- \rightarrow t \bar{t} h$  \\
at linear colliders \footnote{Supported by
National Natural Science Foundation of China.}} \vspace{3mm}
\author{{ You Yu$^{2}$, Ma Wen-Gan$^{1,2}$, Chen Hui$^{2}$, Zhang Ren-You$^{2}$, Sun Yan-Bin$^{2}$,
and Hou Hong-Sheng$^{2}$}\\
{\small $^{1}$ CCAST (World Laboratory), P.O.Box 8730, Beijing
100080, P.R.China}\\
{\small $^{2}$ Department of Modern Physics, University of Science and Technology}\\
{\small of China (USTC), Hefei, Anhui 230027, P.R.China}}
\date{}
\maketitle \vskip 12mm

\begin{abstract}
We calculate the ${\cal O}(\alpha_{{\rm ew}})$ electroweak
radiative corrections to $e^+ e^- \rightarrow t \bar{t} h$ at a
electron-positron linear collider (LC) in the standard model. We
analyze the dependence of the ${\cal O}(\alpha_{{\rm ew}})$
corrections on the Higgs boson mass $m_{h}$ and colliding energy
$\sqrt{s}$, and find that the corrections significantly decrease
or increase the Born cross section depending on the colliding
energy. The numerical results show that the ${\cal O}(\alpha_{{\rm
ew}})$ relative correction is strongly related to the Higgs boson
mass when $\sqrt{s}=500~{\rm GeV}$, and for $m_h = 150~{\rm GeV}$
the relative correction ranges from $-31.3\%$ to $2.3\%$ as the
increment of the colliding energy from 500 GeV to 2 TeV.
\end{abstract}


\vfill \eject \baselineskip=0.36in
\renewcommand{\theequation}{\arabic{section}.\arabic{equation}}
\renewcommand{\thesection}{\Roman{section}}
\newcommand{\nb}{\nonumber}
\makeatletter      
\@addtoreset{equation}{section}
\makeatother       

\section{Introduction}

\par
The main goal of the most experimental programs at present and
future high-energy colliders is to search for Higgs boson, which
is believed to be responsible for the breaking of the electroweak
symmetry and the generation of masses for the fundamental
particles in the standard model (SM) \cite{int higgs1,int higgs2}.
The fundamental particles acquire masses via the interactions with
the ground state Higgs field in the SM Higgs mechanism. However,
until now the Higgs boson hasn't yet been directly explored
experimentally, except that LEP2 experiments provided a lower bound
of 114.4 GeV for the SM Higgs boson mass at the $95\%$ confidence
level \cite{Lep2}.

\par
As we know that the Higgs search strategies depend largely on the
suspected value of Higgs mass. Actually, it is most difficult to
probe Higgs boson experimentally in the intermediate mass region
($m_h \sim 100-200~ {\rm GeV}$). In this mass region, the
production mechanism with Higgs boson radiated from either a gauge
boson or a fermion, is an important Higgs boson discovery channel.
At $e^+e^-$ linear colliders and hadron colliders, the Higgs boson
is searched via Bjorken process $e^+e^-(p\bar{p},pp) \rightarrow f
\bar{f} h$, an intermediate Higgs boson is produced associated
with a $f\bar{f}$ pair. The coupling strength of the fermion-Higgs
Yukawa coupling $f-\bar{f}-h$ is proportional to the fermion mass,
i.e., $g_{f \bar{f} h} = m_f/v$, where $v = (\sqrt{2}G_{F})^{-1/2}
\simeq 246~{\rm GeV}$ is the vacuum expectation value of the Higgs
boson. Then we can see that the top quark Yukawa coupling $g_{t
\bar{t} h}$ is the largest one among all the fermion-Higgs
couplings, e.g., $g_{t \bar{t} h}^2 \simeq 0.5$ to be compared for
example with $g_{b \bar{b} h}^2 \simeq 4 \times 10^{-4}$.
Therefore, the Higgs boson production via the process $e^+e^-
\rightarrow t \bar{t} h$ is strongly enhanced by the top quark
Yukawa coupling, and it can also be a basic mechanism for
measuring the top quark Yukawa coupling.

\par
Recently, a lot of efforts have been invested in improving the
precise QCD theoretical corrections to the processes $p\bar{p}/pp
\rightarrow t\bar{t}h + X$ \cite{ppbar, ppbar1, ppbar2, ppbar3, eq1}.
At a LC the cross section for $e^{+}e^{-} \rightarrow t
\bar{t} h$ is small, about $1~{\rm fb}$ for $\sqrt{s} = 500~{\rm GeV}$ and
$m_h=100~ {\rm GeV}$ \cite{ee, ee1, ee2}. But it has a distinctive
experimental signature and can potentially be used to measure the
top quark Yukawa coupling in the intermediate Higgs mass region at
a LC with very high luminosity. In Ref. \cite{ee1}, S. Dawson and
L. Reina found that the NLO QCD corrections increase the Born
cross section by a factor of roughly 1.5 for $e^+e^- \rightarrow
t \bar{t} h$ at $\sqrt{s} = 500~ {\rm GeV}$ and $m_{h}=100~ {\rm GeV}$. But at
$\sqrt{s}=1~ {\rm TeV}$, the corrections decrease the Born cross section
and are relative small. These works show that the evaluation of
radiative corrections is a crucial task for all accurate
measurements of this process.

\par
In this paper we present the calculations of the full ${\cal
O}(\alpha_{{\rm ew}})$ electroweak corrections to $e^+e^-
\rightarrow t\bar{t}h$ in the SM. In section 2, we present our
calculations of the full ${\cal O}(\alpha_{{\rm ew}})$
electroweak radiative corrections. The
numerical results and discussions are presented in section 3.
Finally, a short summary is given.

\section{Calculations}

\par
In this paper we use the t'Hooft-Feynman gauge. In the
calculations of loop diagrams we adopt the definitions of one-loop
integral functions of Ref. \cite{s14}. The Feynman diagrams and
the relevant amplitudes are created by {\it FeynArts} 3 \cite{FA3}
automatically, and the Feynman amplitudes are subsequently reduced
by {\it FORM} \cite{FORM}. The numerical calculations of integral
functions are implemented by using Fortran programs, in which the
5-point loop integrals are evaluated by using the approach
presented in Ref. \cite{fivep}.

\begin{figure}[htp]
\centering
\includegraphics*[65,570][530,685]{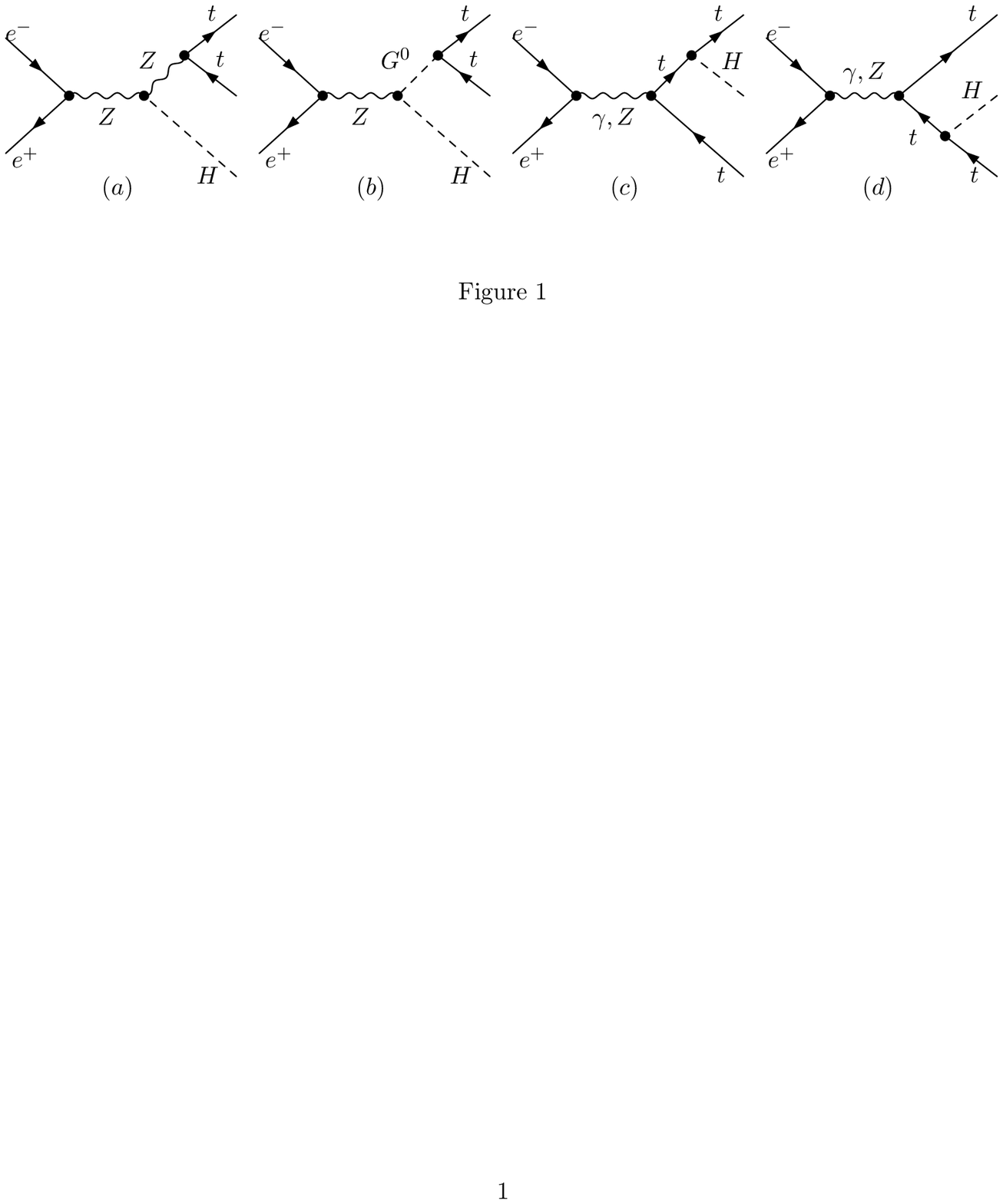}
\center{{\bf Figure 1} The tree-level Feynman diagrams for the
$e^+e^- \rightarrow t\bar{t}h$ process.}
\end{figure}

\par
The process $e^+e^- \rightarrow t\bar{t} h$ at the lowest level occurs
through the Feynman diagrams of Fig.1. There are two kinds of
Feynman diagrams to the $e^+e^- \to t\bar{t}h$ process at the tree-level.
The first kind includes diagrams with Higgsstrahlungs from
top or anti-top final states and the top Yukawa coupling is thus
involved. The second kind consists of the diagrams with a Higgs
boson radiated from a $Z^0$-exchange s-channel or via
$Z^0-G^0-h$-interaction, and is independent of the top Yukawa
coupling. We have checked that our numerical Born cross section
is in good agreement with that in Ref. \cite{ee1, ee3, born}.

\par
The ${\cal O}(\alpha_{{\rm ew}})$ (one-loop level) virtual corrections to the process
\begin{eqnarray}
e^+(p_1)+e^-(p_2) \rightarrow t(k_1)+\bar{t}(k_2)+h(k_3)
\end{eqnarray}
can be expressed as
\begin{eqnarray}
\sigma_{{\rm virtual}} = \sigma_0 \delta_{{\rm virtual}} =
\frac{N_c}{2|\vec{k}_1|\sqrt{s}}\int {\rm d} \Phi_3\overline{\sum\limits_{{\rm spin}}}
{\rm Re}\left( {\cal M}_0 {\cal M}_{{\rm virtual}}^* \right),
\end{eqnarray}
where $\sigma_0$ and ${\cal M}_0$ are the Born cross section and
amplitude for $e^+e^- \to t\bar{t}h$, respectively,
${\rm d} \Phi_3$ is the three-body phase space element, the bar
over summation recalls averaging over initial spins, and ${\cal
M}_{{\rm virtual}}$ is the amplitude of one-loop Feynman diagrams
and the corresponding counter-terms. Due to the fact that the
electron-Higgs(Goldstone) Yukawa coupling is proportional to the
electron mass, we do not consider the contributions of the
one-loop diagrams which involve $\bar{e}-e-h(G^0)$ vertex to the
amplitude ${\cal M}_{{\rm virtual}}$. Therefore, the ${\cal O}(\alpha_{{\rm ew}})$
virtual corrections involve 975 loop diagrams which can be
classified into self-energy (376), vertex (425), box (145) and
pentagon (29) diagrams. As a representative selection, the
pentagon diagrams are given in Fig.2.

\begin{figure}[htp]
\centering
\includegraphics*[85,315][490,710]{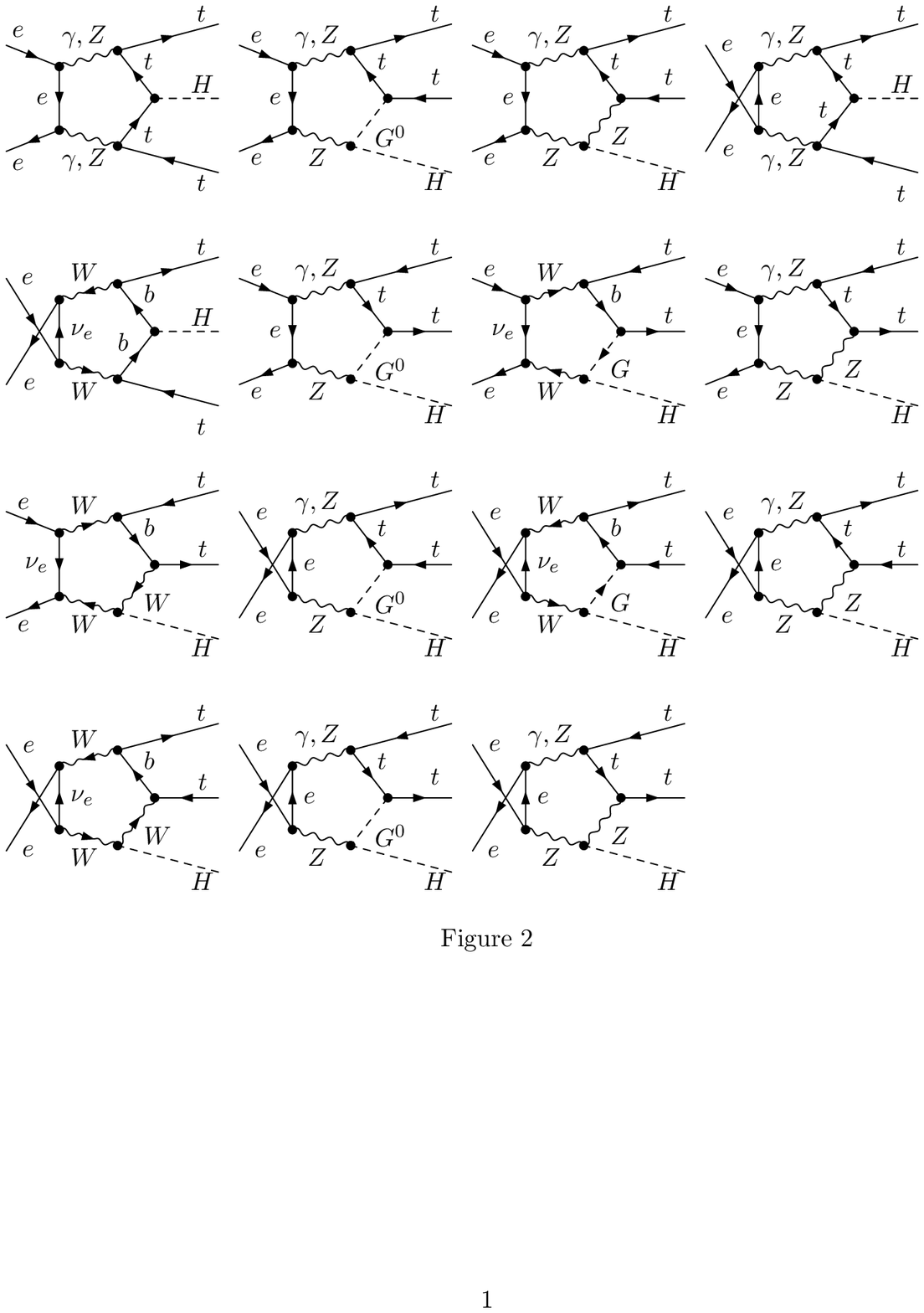}
\center{{\bf Figure 2} The pentagon diagrams for the $e^+e^-
\rightarrow t\bar{t}h$ process.}
\end{figure}

\par
The self-energy and vertex one-loop diagrams contain both
ultraviolet (UV) and infrared (IR) divergences, while all the box
and pentagon diagrams are ultraviolet finite and only contain IR
divergences. To regularize the UV divergences in loop integrals,
we adopt the dimensional regularization scheme \cite{DR} in which
the dimensions of spinor and spacetime manifolds are extended to
$D = 4 - 2 \epsilon$. In this paper, we adopt the complete
on-mass-shell (COMS) renormalization scheme \cite{COMS}, in which
the electric charge of electron $e$ and the physical masses $m_W$,
$m_Z$, $m_h$, $m_t$, $m_e$ are chosen to be the renormalized
parameters. The definitions and the explicit expressions of these
renormalization constants can be found in Ref. \cite{COMS}. As we
expect, the UV divergence contributed by the one-loop diagrams can
be cancelled by that contributed by the counterterms exactly.
Therefore, we get a UV finite cross section including ${\cal O}(\alpha_{{\rm ew}})$
virtual radiative corrections.

\par
The IR divergence in the process $e^+e^- \rightarrow t\bar{t}h$ is
originated from virtual photonic corrections. It can be exactly
cancelled by including the real photonic bremsstrahlung
corrections to this process in the soft photon limit. The real
photon emission process
\begin{eqnarray}
\label{real photon emission}
 e^+(p_1)+e^-(p_2) \rightarrow t(k_1)+\bar{t}(k_2)+h(k_3)+\gamma(k),
\end{eqnarray}
where the real photon radiates from the initial electron(positron)
and the final top(anti-top) quark, can have either soft or
collinear nature. The collinear singularity is regularized by
keeping electron mass. In order to isolate the soft photon
emission singularity in the real photon emission process, we use
the general phase-space-slicing method \cite{PSS}. The
bremsstrahlung phase space is divided into singular and
non-singular regions, and the cross section of the real photon
emission process ($\ref{real photon emission}$) is decomposed into
soft and hard terms
\begin{equation}
\sigma_{{\rm real}}=\sigma_{{\rm soft}}+\sigma_{{\rm hard}}=
\sigma_0(\delta_{{\rm soft}}+\delta_{{\rm hard}}).
\end{equation}
For soft photons, $k_0<\Delta E$, we neglect the momenta of the
radiated photons everywhere but in the singular propagators. By using
the soft photon approximation, we find the contribution of the
soft photon emission process is \cite{COMS,Velt}
\begin{eqnarray}
\label{approsoft}
{\rm d} \sigma_{{\rm soft}} = -{\rm d} \sigma_0 \frac{\alpha_{{\rm ew}}}{2 \pi^2}
 \int_{|\vec{k}| \leq \Delta E}\frac{{\rm d}^3 k}{2 k_0} \left[
 \frac{Q_e p_1}{p_1\cdot k}-\frac{Q_e p_2}{p_2\cdot k}
 -\frac{Q_t k_1}{k_1\cdot k}+\frac{Q_t k_2}{k_2\cdot k} \right]^2,
\end{eqnarray}
in which $\Delta E$ is the energy cutoff of the soft photon and
$k_0 \leq \Delta E \ll \sqrt{s}$, $Q_e=1$ and $Q_t=2/3$ are the
electric charges of the positron and top quark. $k_0 =
\sqrt{|\vec{k}|^2+\mu^2}$ is the photon energy. The integral over
the soft photon phase space have been implemented, therefore, we obtain the
analytical result of the soft corrections to
$e^+e^- \rightarrow t\bar{t} h$ which can be found in Refs. \cite{COMS} and
\cite{Velt}.
We checked numerically the cancellation of IR divergencies and
verified the contribution of these soft photonic bremsstrahlung
corrections leads to a IR finite cross section which is
independent of the infinitesimal photon mass $\mu$. The hard
photon emission cross section $\sigma_{{\rm hard}}$, with the
radiated photon energy being larger than $\Delta E$, is calculated
by using Monte Carlo method.

\par
Finally the UV and IR finite total cross section including the
full ${\cal O}(\alpha_{{\rm ew}})$ electroweak corrections reads
\begin{equation}\label{cs}
\sigma  = \sigma_0 + \Delta \sigma
=
\sigma_0 +
\sigma_{{\rm virtual}} +
\sigma_{{\rm real}}
= \sigma_0(1 + \delta),
\end{equation}
where
$\delta =
\delta_{{\rm virtual}} +
\delta_{{\rm soft}} +
\delta_{{\rm hard}}
$
is the full electroweak relative correction of the order ${\cal O}(\alpha_{{\rm ew}})$.

\vskip 10mm
\section{Numerical results and discussions}

\par
In our numerical calculations, we set \cite{pdg}: $\alpha_{{\rm
ew}}(0)^{-1} = 137.0359895$, $m_W = 80.423~{\rm GeV}$, $m_Z =
91.188~{\rm GeV}$, $m_e = 0.511~{\rm MeV}$, $m_{\mu} = 105.7~{\rm
MeV}$, $m_{\tau} = 1.777~ {\rm GeV}$, $m_u = 66~{\rm MeV}$, $m_c =
1.35~{\rm GeV}$, $m_t = 174.3~{\rm GeV}$, $m_d = 66~{\rm MeV}$,
$m_s = 150~{\rm MeV}$, and $m_b = 4.3~{\rm GeV}$. The
renormalization scale is taken to be $Q = 2m_t + m_h$. Here we use
the effective values of the light quark masses ($m_u$ and $m_d$)
which can reproduce the hadron contribution to the shift in the
fine structure constant $\alpha_{{\rm ew}}(m_Z)$ \cite{jeger}.

\begin{figure}[htp]
\centering
\scalebox{1}{\includegraphics*[20,30][469,337]{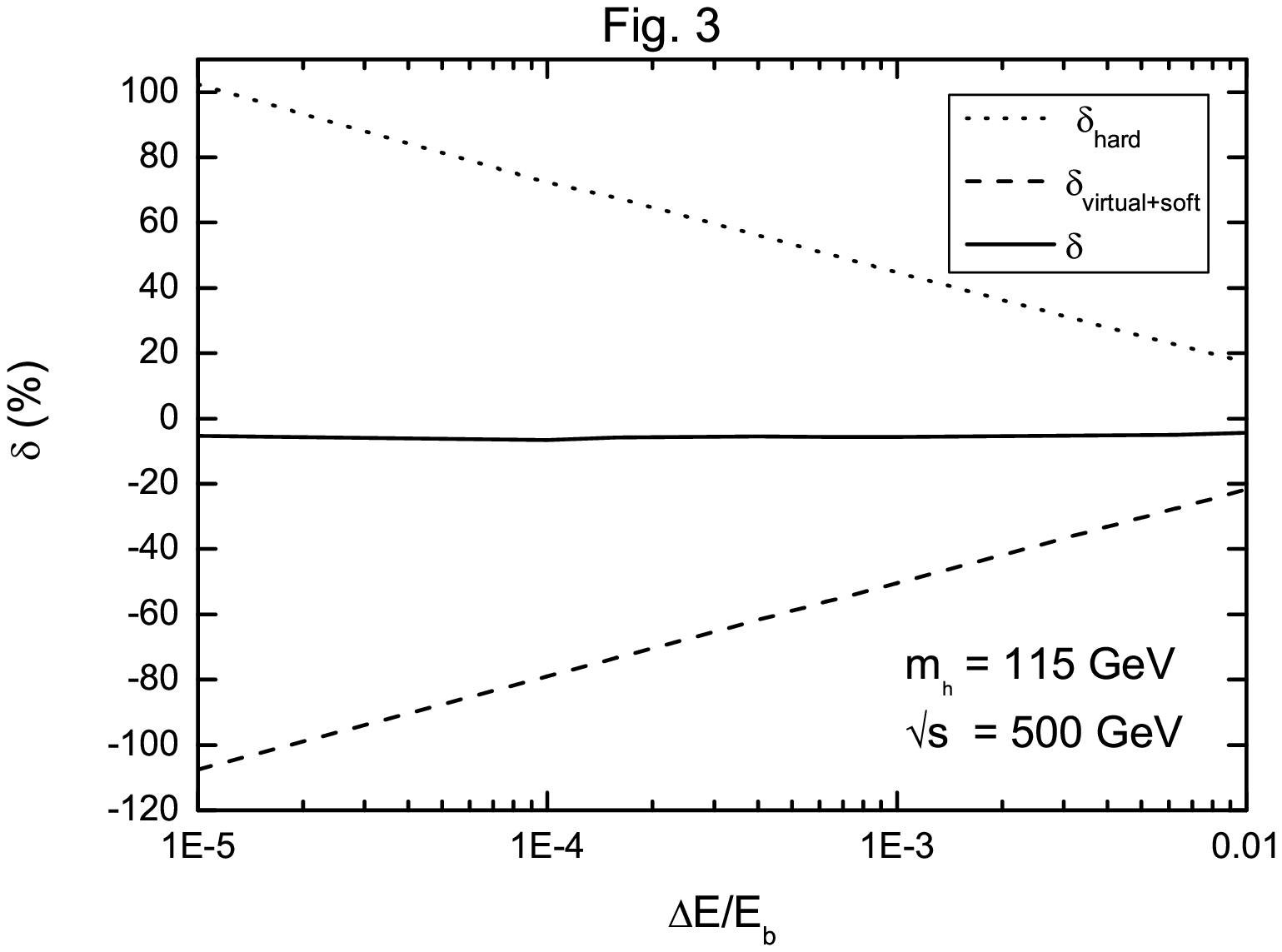}}
\center{{\bf Figure 3} The relative corrections of the ${\cal
O}(\alpha_{{\rm ew}})$ order contributions to the $e^+ e^-
\rightarrow t \bar{t} h$ cross section. The relative corrections
of $\delta_{virtual+soft}$ and $\delta_{hard}$, and their sum are
shown as the functions of the soft photon energy cutoff $\Delta
E/E_b$.}
\end{figure}

\par
In Fig.3 we show the dependence of the ${\cal O}(\alpha_{{\rm
ew}})$ relative correction to $e^+ e^- \rightarrow t \bar{t} h$ on
the soft cutoff $\Delta E/E_b$, assuming $m_{h} = 115~ {{\rm
GeV}}$ and $\sqrt{s} = 500~ {\rm GeV}$. As shown in this figure,
both $\delta_{{\rm virtual}+{\rm soft}}(=\delta_{\rm
virtual}+\delta_{\rm soft})$ and $\delta_{{\rm hard}}$ depend on
the soft cutoff $\Delta E/E_b$, but the full ${\cal
O}(\alpha_{{\rm ew}})$ electroweak relative correction $\delta$ is
cutoff independent within the range of statistical errors as
expected. In the following calculations, the soft cutoff $\Delta
E/E_b$ is fixed to be $10^{-3}$.

\begin{figure}[htp]
\centering
\scalebox{1}{\includegraphics*[30,30][450,335]{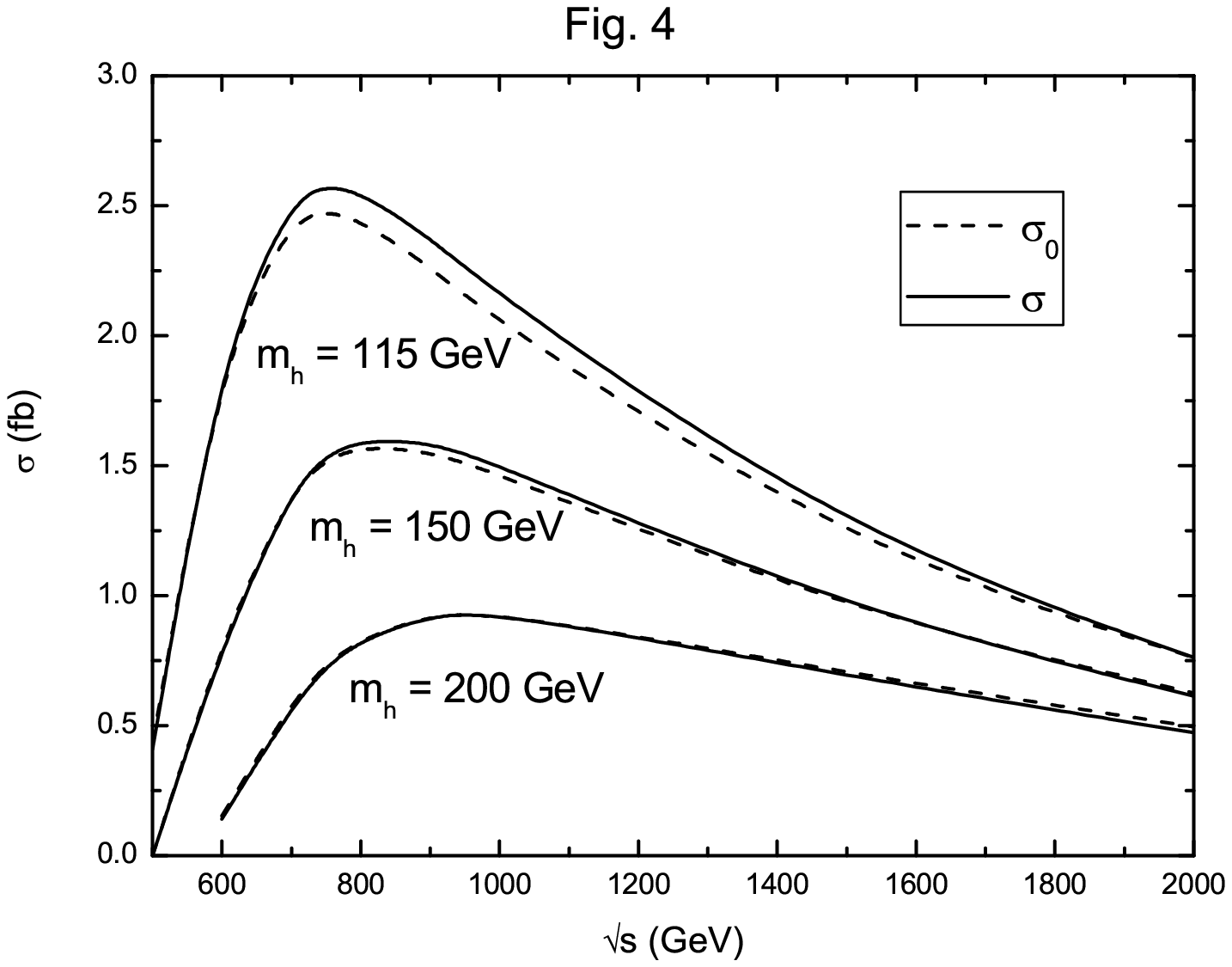}}
\center{{\bf Figure 4} The Born and one-loop level cross sections
for the process $e^+ e^- \rightarrow t \bar{t} h$ as functions of
the $e^+e^-$ colliding energy $\sqrt{s}$.}
\end{figure}

\par
In Fig.4 we present the Born cross section $\sigma_0$ and the
one-loop level cross section $\sigma$ which include the full
${\cal O}(\alpha_{{\rm ew}})$ electroweak corrections, as
functions of the $e^+ e^-$ colliding energy $\sqrt{s}$ for the
Higgs boson mass $m_h = 115~{\rm GeV}$, 150 GeV and 200 GeV,
respectively. The colliding energy range ($500~ {\rm GeV} \sim 2~
{\rm TeV}$) in the figure is accessible at future linear
colliders, such as, TESLA \cite{TESLA} ($\sqrt{s} = 500~ {\rm
GeV}$), NLC \cite{NLC} ($\sqrt{s} = 500~ {\rm GeV}$), JLC
\cite{JLC} ($\sqrt{s} = 500~ {\rm GeV}$,) and CERN CLIC
\cite{CLIC} ($1~ {\rm TeV} < \sqrt{s} < 5~ {\rm TeV}$). From this
figure we can see that both curves of $\sigma_0$ and $\sigma$ for
$m_h = 115~ {\rm GeV}$ reach maximal values at the position of
$\sqrt{s} \sim 750~ {\rm GeV}$, where the correction $\Delta
\sigma$ can reach $0.09~{\rm fb}$. For $m_h = 150~{\rm GeV}$ and
200 GeV, both $\sigma_0$ and $\sigma$ have their maximal values at
$\sqrt{s} \sim 850~{\rm GeV}$ and $\sqrt{s} \sim 950~{\rm GeV}$,
respectively. The correction $\Delta \sigma$ decreases as the
increment of the Higgs boson mass $m_h$.

\begin{figure}[htp]
\centering
\scalebox{1}{\includegraphics*[30,30][459,334]{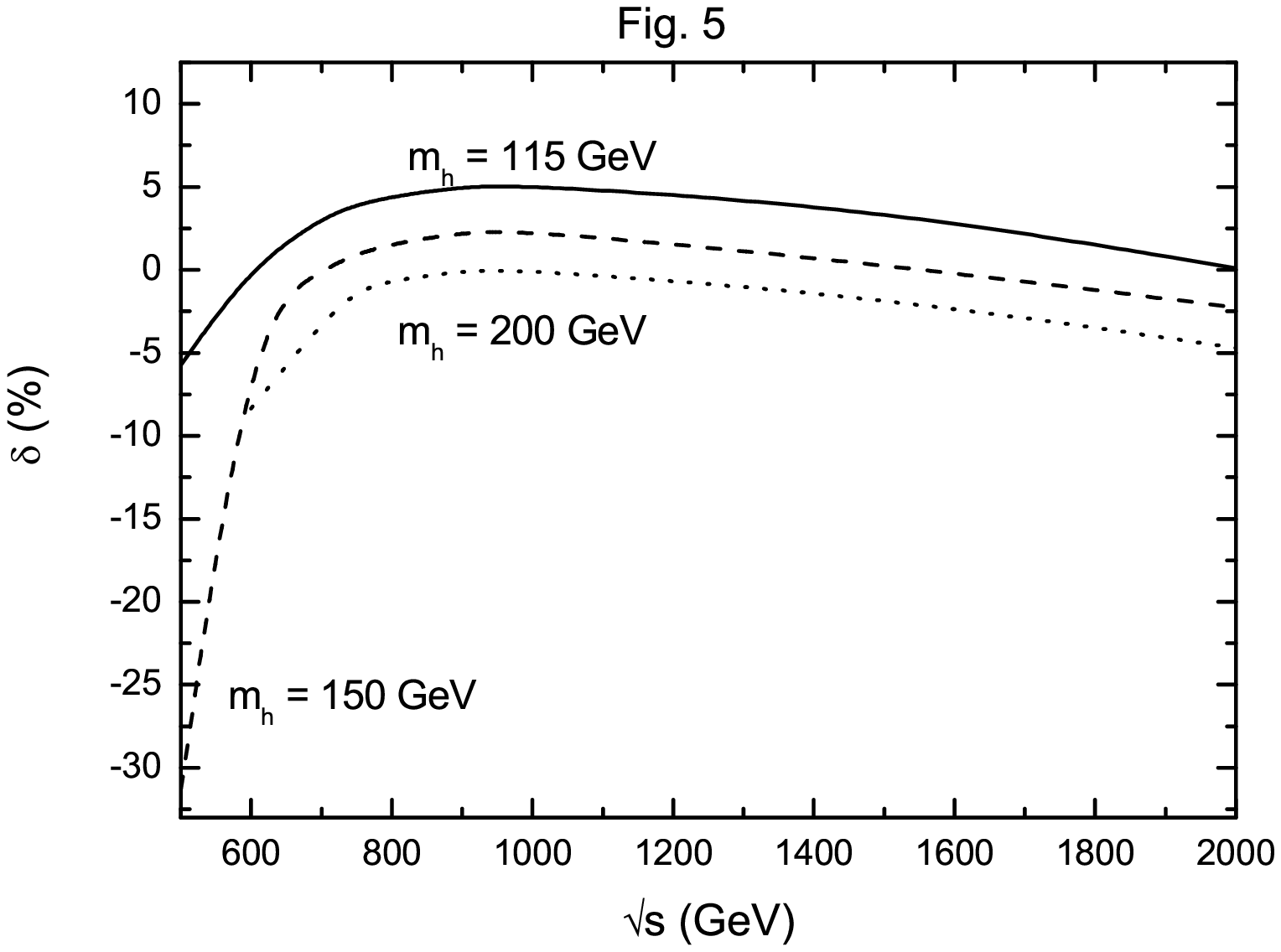}}
\center{{\bf Figure 5} The dependence of the ${\cal
O}(\alpha_{{\rm ew}})$ relative correction to $e^+ e^- \rightarrow
t \bar{t} h$ on the $e^+ e^-$ colliding energy $\sqrt{s}$.}
\end{figure}

\par
The dependence of the ${\cal O}(\alpha_{{\rm ew}})$ relative
correction to $e^+ e^- \rightarrow t \bar{t} h$ on the colliding
energy $\sqrt{s}$ of a LC is displayed in Fig.5. For $m_h =
115~{\rm GeV}$, the ${\cal O}(\alpha_{{\rm ew}})$ corrections
suppress the Born cross section in the energy region of $\sqrt{s}
< 600~{\rm GeV}$, while enhance the Born cross section in the
region of $\sqrt{s} > 600~{\rm GeV}$. The relative corrections can
reach about $-5.8\%$, $-31.3\%$ and $-8.3\%$ at $\sqrt{s} = 500$,
500 and 600 GeV for $m_h = 115$, 150 and 200 GeV, respectively.
For $m_h = 150~ {\rm GeV}$, the relative correction ranges from
$-31.3\%$ to $2.3\%$ as $\sqrt{s}$ running from $500~{\rm GeV}$ to
$2~{\rm TeV}$. The large correction for $\sqrt{s}=500~GeV$ and
$m_h \sim 150~GeV$ shown in this figure comes from a threshold
effect which diverges at the threshold.

\begin{figure}[htp]
\centering
\scalebox{1}{\includegraphics*[30,30][467,344]{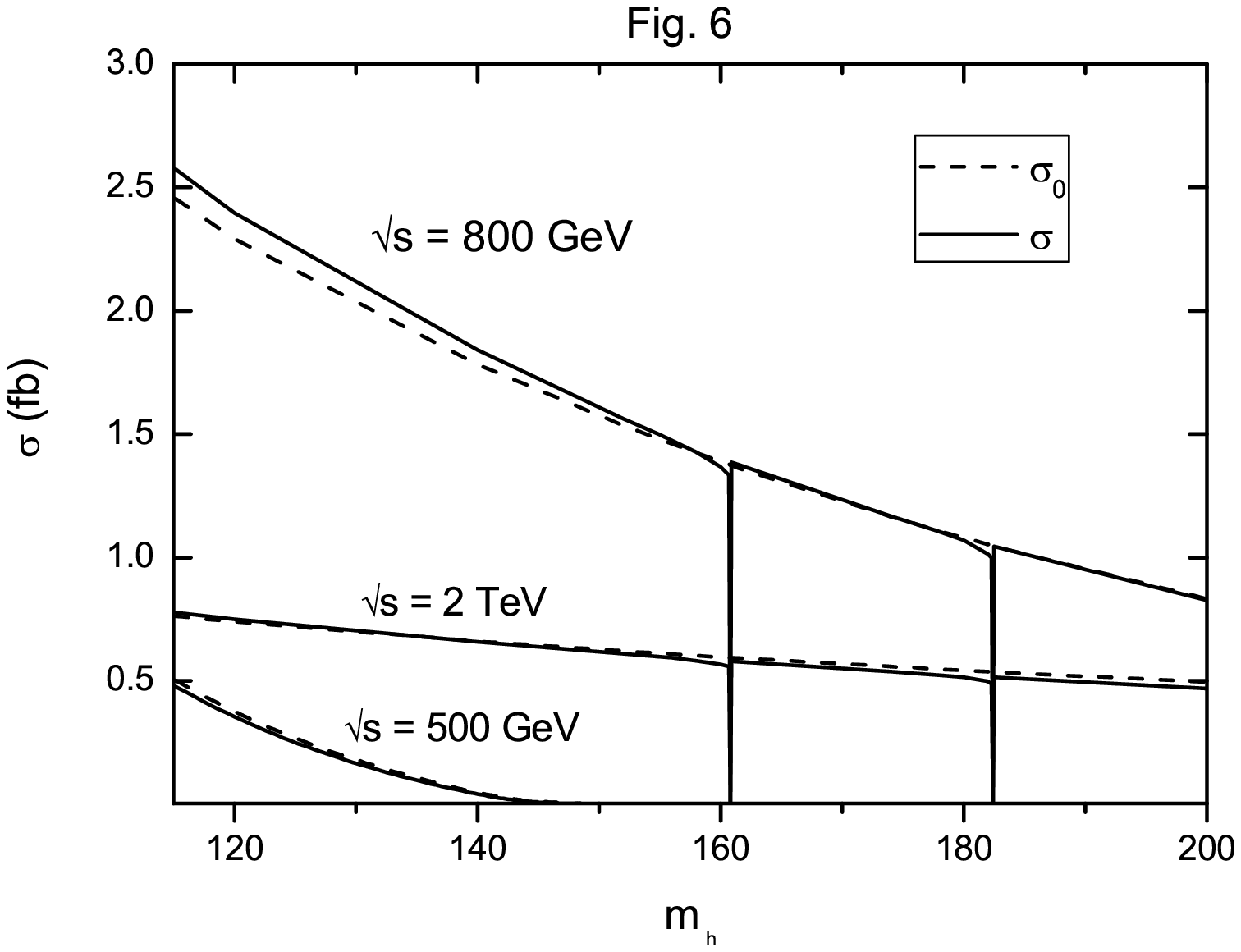}}
\center{{\bf Figure 6} The Born and one-loop level cross sections
for the process $e^+ e^- \rightarrow t \bar{t} h$ as functions of
the Higgs boson mass $m_h$.}
\end{figure}

\par
In Fig.6 we depict the Born cross section $\sigma_0$ and the
one-loop level cross section $\sigma$ as functions of the Higgs
boson mass $m_h$. As shown in this figure, Both the cross sections
$\sigma_0$ and $\sigma$ decrease with the increment of the Higgs
boson mass, and the cross sections $\sigma_0$, $\sigma$ and the
one-loop level correction $\Delta \sigma = \sigma - \sigma_0$ at
$\sqrt{s} = 800~{\rm GeV}$ are larger than those at $\sqrt{s} =
500~{\rm GeV}$ and $\sqrt{s}=2~{\rm TeV}$.

\begin{figure}[htp]
\centering
\scalebox{1}{\includegraphics*[30,30][467,342]{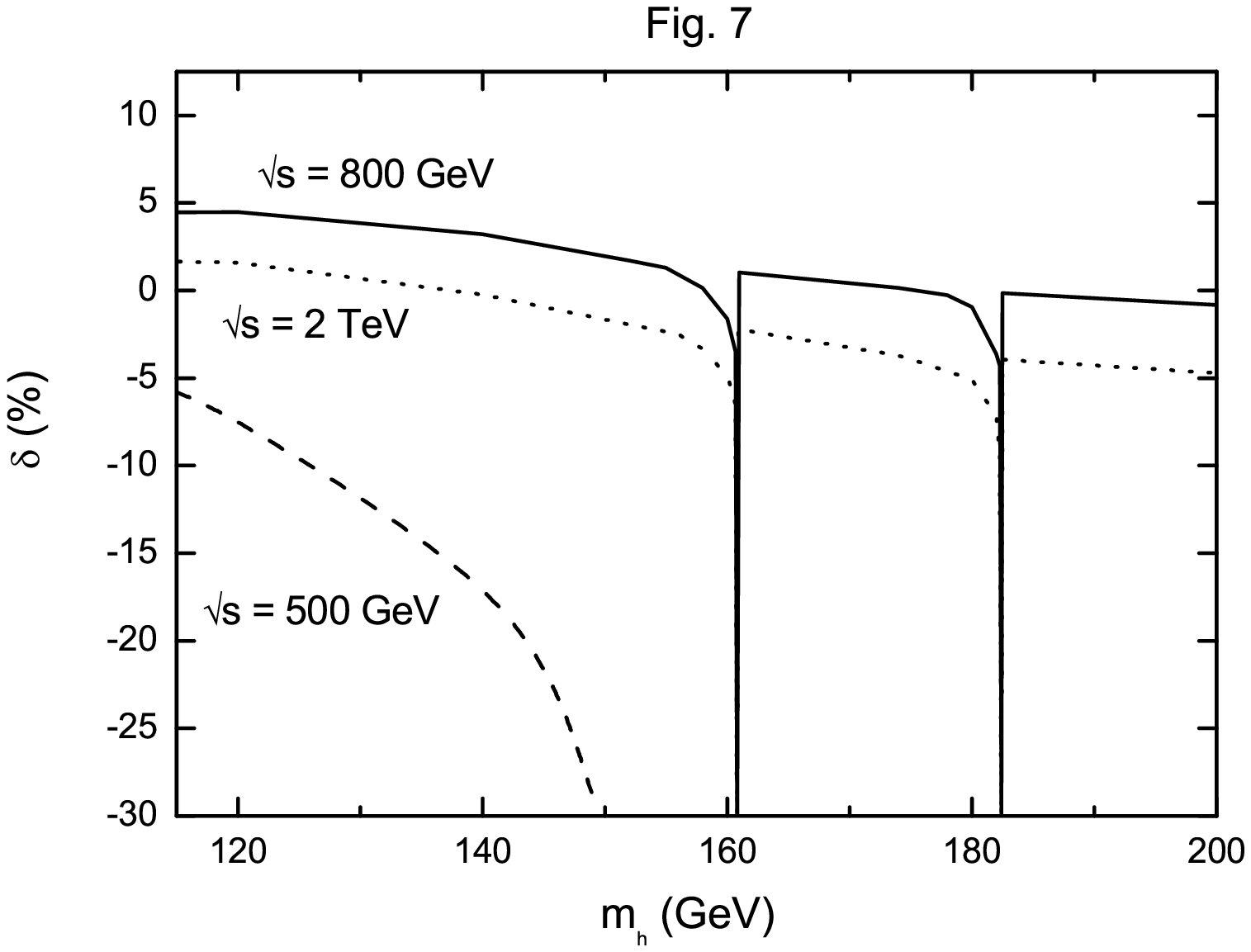}}
\center{{\bf Figure 7} The ${\cal O}(\alpha_{{\rm ew}})$ relative
correction to the process $e^+ e^- \rightarrow t \bar{t} h$ as a
function of the Higgs boson mass $m_h$.}
\end{figure}

\par
In Fig.7 we plot the ${\cal O}(\alpha_{{\rm ew}})$ relative
correction $\delta$ to $e^+ e^- \rightarrow t \bar{t} h$ as a
function of $m_h$. For $\sqrt{s} = 500~{\rm GeV}$, the relative
correction decreases from $-5.8\%$ at $m_h = 115~{\rm GeV}$ to
$-31.3\%$ at $m_h = 150~{\rm GeV}$. For $\sqrt{s} = 800~{\rm GeV}$
and 2 TeV, the relative corrections decrease from $4.4\%$ and
$1.5\%$ to $-0.8\%$ and $-4.7\%$ as the increment of $m_h$ from
115 GeV to 200 GeV, respectively. The ${\cal O}(\alpha_{{\rm
ew}})$ electroweak relative corrections are not very sensitive to
the Higgs boson mass in the range of $115~{\rm GeV} < m_h <
200~{\rm GeV}$ when $\sqrt{s} = 800~ {\rm GeV}$ and 2 TeV, but
strongly depend on the Higgs boson mass when $\sqrt{s} = 500~{\rm
GeV}$. In both Fig.6 and Fig.7, we can see that each of the two
curves of the total cross sections including the ${\cal
O}(\alpha_{{\rm ew}})$ corrections at $\sqrt{s} = 800~{\rm GeV}$
and $2~{\rm TeV}$, has two spikes at the vicinities of $m_h = 2
m_W$ and $m_h = 2 m_Z$, due to the threshold effects.
\par
After submitting this manuscript, we acknowledged another two
papers appeared on this subject \cite{ref1,ref2}. The
representative comparison with the calculation of Ref.\cite{ref2}
is shown in Table 2 of \cite{ref2}(There we use the current mass
values for $m_u$ and $m_d$.). It shows that most of our numerical
results of one-loop electroweak corrected cross sections agree
with theirs within estimated error, but there are some
discrepancies at the energy $\sqrt{s}=2~TeV$.

\section{ Summary}
\par
In this paper we calculate the full electroweak one-loop level
radiative corrections to the process $e^+ e^- \rightarrow t
\bar{t} h$ at a electron-positron LC in the standard model. We
analyze the dependence of the electroweak radiative corrections on
the Higgs boson mass $m_{h}$ and colliding energy $\sqrt{s}$, and
find that the corrections increase or decrease the Born cross
section in the Higgs boson mass range $115~{\rm GeV} < m_h <
200~{\rm GeV}$, depending on the colliding energy. The numerical
results show that the ${\cal O}(\alpha_{{\rm ew}})$ electroweak
relative corrections can reach $-31.3\%$, $4.4\%$ and $-4.7\%$ at
$\sqrt{s} = 500~{\rm GeV}$, 800 GeV and 2 TeV, respectively. We
also find that the full electroweak relative correction of the
order ${\cal O}(\alpha_{{\rm ew}})$ is strongly related to the
Higgs boson mass when $\sqrt{s}= 500~ {\rm GeV}$, and for $m_h =
150~ {\rm GeV}$ the relative correction ranges from $-31.3\%$ to
$2.3\%$ as the colliding energy increasing from 500 GeV to 2 TeV.

\vskip 10mm \noindent{\large\bf Acknowledgments:} This work was
supported in part by the National Natural Science Foundation of
China and a grant from the University of Science and Technology of
China.

\vskip 10mm
\begin{flushleft} {\bf Figure Captions} \end{flushleft}
\par
{\bf Figure 1} The tree-level Feynman diagrams for the $e^+e^- \rightarrow
t\bar{t}h$ process.
\par
{\bf Figure 2} The pentagon diagrams for the $e^+e^- \rightarrow t\bar{t}h$ process.
\par
{\bf Figure 3} The relative corrections of the ${\cal
O}(\alpha_{{\rm ew}})$ order contributions to the $e^+ e^-
\rightarrow t \bar{t} h$ cross section. The relative corrections
of $\delta_{virtual+soft}$ and $\delta_{hard}$, and their sum are
shown as a function of the soft photon energy cutoff $\Delta
E/E_b$.

\par
{\bf Figure 4} The Born and one-loop level cross sections for the process
$e^+ e^- \rightarrow t \bar{t} h$ as functions of
the $e^+e^-$ colliding energy $\sqrt{s}$.

\par
{\bf Figure 5} The dependence of the ${\cal O}(\alpha_{{\rm ew}})$
relative correction to $e^+ e^- \rightarrow t \bar{t} h$ on the
$e^+ e^-$ colliding energy $\sqrt{s}$.

\par
{\bf Figure 6}  The Born and one-loop level cross sections for the process
$e^+ e^- \rightarrow t \bar{t} h$ as functions of
the Higgs boson mass $m_h$.

\par
{\bf Figure 7} The ${\cal O}(\alpha_{{\rm ew}})$ relative correction to the
process $e^+ e^- \rightarrow t \bar{t} h$ as a function of the Higgs boson mass $m_h$.
\end{document}